\documentclass[prb,aps,twocolumn,showpacs,nobibnotes,epsf]{revtex4}

\usepackage{graphicx}
\usepackage{dcolumn}
\usepackage{bm}
\usepackage{SIunits}
\usepackage{tabularx}

\begin{document}
\title{Pressure effects on superconducting properties of single-crystalline Co doped NaFeAs }
\author{A. F. Wang, Z. J. Xiang, J. J. Ying, Y. J. Yan, P. Cheng, G. J. Ye, X. G. Luo and X. H. Chen}
\altaffiliation{Corresponding author} \email{chenxh@ustc.edu.cn}
\affiliation{Hefei National Laboratory for Physical Science at Microscale and Department of
Physics, University of Science and Technology of China, Hefei, Anhui 230026, People's Republic
of China\\}

\begin{abstract}
Resistivity and magnetic susceptibility measurements under external
pressure were performed on single-crystals NaFe$_{1-x}$Co$_x$As
(x=0, 0.01, 0.028, 0.075, 0.109). The maximum $T_{\rm c}$ enhanced
by pressure in both underdoped and optimally doped
NaFe$_{1-x}$Co$_x$As is the same, as high as 31 K. The overdoped
sample with x = 0.075 also shows a positive pressure effect on $T_{\rm c}$, and
an enhancement of $T_{\rm c}$ by 13 K is achieved under pressure of 2.3
GPa. All the superconducting
samples show large positive pressure coefficient on
superconductivity, being different from
Ba(Fe$_{1-x}$Co$_x$)$_2$As$_2$. However, the superconductivity
cannot be induced by pressure in heavily overdoped
non-superconducting NaFe$_{0.891}$Co$_{0.109}$As. These results
provide evidence for that the electronic structure is much different
between superconducting and heavily overdoped non-superconducting
NaFe$_{1-x}$Co$_x$As, being consistent with the observation by
angle-resolved photoemission spectroscopy.
\end{abstract}


\vskip 300 pt

\maketitle

\section{INTRODUCTION}

Extensive experimental and theoretical efforts have been made to
study the iron based superconductors since the discovery of
superconductivity in F doped LaOFeAs.\cite{Hosono} Most of the
parent compounds of the iron based superconductors undergo
structural and spin density wave (SDW) transitions. With doping or
applying high pressure, both the structural and SDW transitions are
suppressed and superconductivity emerges. The so called "111"-type
iron arsenide compound with the PbFCl structure, including
LiFeAs\cite{WangXC} and NaFeAs\cite{Parker1}, has been regarded as a
unique family which is superconducting without purposely doping or
applying pressure. Although no long range antiferromagnetic order
has been observed in LiFeAs\cite{Borisenko}, NaFeAs is reported to
undergo three successive phase transitions at around 52, 41, and 23
K, which correspond to structural, magnetic, and superconducting
transition, respectively.\cite{ChenGF} Although the resistivity of
NaFeAs drops to zero at about 10K, it has been pointed out that the
superconductivity is filamentary rather than a bulk
phenomenon.\cite{DaiPC, WangAF} With substitution of Co on Fe site, both
magnetism and the structural distortion are suppressed, and bulk
superconductivity with zero resistivity up to 20 K can be
achieved.\cite{Parker2, Wright} Full shielding fraction and large
specific heat jump can be observed in single-crystalline optimally
doped NaFe$_{0.972}$Co$_{0.028}$As samples.\cite{WangAF}

Applying pressure has been proved to be an effective method to
enhance the superconductivity transition temperature in many types
of iron arsenide superconductors. It was revealed that the $T_{\rm
c}$ of  F-doped LaOFeAs was enhanced up to 43 K soon after the
discovery of superconductivity in this system.\cite{Takahashi} In
tetragonal FeSe, $T_{\rm c}$ increases from 8.5 K at ambient pressure
to about 37 K under P = 8.9 GPa, which is the largest pressure effect
reported in iron based superconductors so far.\cite{FeSepressure}
Pressure effects in electron doped 122-system
Ba(Fe$_{1-x}$Co$_x$)$_2$As$_2$ with different doping levels have
been thoroughly studied. Applying pressure dramatically enhances
$T_{\rm c}$ in the underdoped regime, whereas, the effect of
pressure on $T_{\rm c}$ is rather small in the optimally doped and
overdoped regimes.\cite{Ahilan, Colombier} For the "111"-type
Fe-pnictides, it is reported that the transition temperature of
LiFeAs is suppressed linearly with pressure\cite{Gooch}, whereas,
$T_{\rm c}$ of Na$_{1-x}$FeAs polycrystal can be enhanced up to 31 K
at about 3 GPa.\cite{JinCQ} This difference is attributed to the
different ionic radius between Li and Na. However, in former
high-pressure study the superconducting transition is rather broad
due to the highly hygroscopic nature of polycrystalline NaFeAs
sample. In order to study the intrinsic properties of this system,
it is of great interest to investigate the combined effect of doping
and pressure on superconducting properties of single-crystal samples. In this
paper, we report the results of resistivity measurements under
hydrostatic pressure for single-crystalline NaFe$_{1-x}$Co$_x$As,
tracking $T_{\rm c}$ as a function of both pressure and doping level
in different regions of the phase diagram. The initial slope of the pressure dependence of $T_{\rm c}$,
(d$T_{\rm c}$/d$P$)$_{\rm P=0}$, is positive in the whole
superconducting doping regime of phase diagram. The value of
pressure coefficient is comparably large among Fe-pnictides, even in
the overdoped region. For the nonsuperconducting extremely overdoped
sample, the pressure effect is negligible. $T_{\rm c}^{\rm offset}$
as high as 31 K, generally consistent with the maximum
$T_{\rm c}$ under pressure in polycrystalline NaFeAs, can be reached
in both underdoped and optimally doped samples. The identical
maximum $T_{\rm c}$ under pressure in different doping regions
indicates that there is a universal maximum transition temperature
of about 31 K in electron-doped NaFeAs, which can be obtain by
applying high pressure or combined effect of pressure and doping.

\section{EXPERIMENTAL DETAILS}

High-quality NaFe$_{1-x}$Co$_x$As single crystals were grown by the
conventional high temperature solution growth method using the NaAs
self-flux technique. Details of the growth procedures were provided
in our previous work.\cite{WangAF} Electrical
resistivity was measured using the ac four-probe method. Pressure
was generated using a Be-Cu pressure cell with a Teflon cup which
was filled with Daphene Oil 7373. The pressure applied in the
resistivity measurement was determined by shift of the
superconducting transition temperature of pure Sn.\cite{ZhangM} The
magnetic susceptibility was measured under pressure up to 6.1 GPa in
a diamond anvil cell (DAC). The pressure transmitting medium was
Daphene Oil 7373 and the pressure was measured at room temperature by
ruby fluorescence spectroscopy. The resistivity measurements were
performed using a Quantum Design physical properties measurement
system (PPMS-9), and the magnetic susceptibility was measured using
a superconducting quantum-interference device magnetometer
(SQUID-MPMS-7T, $Quantum$ $Design$).

\begin{figure}[ht]
\centering
\includegraphics[width=0.49\textwidth]{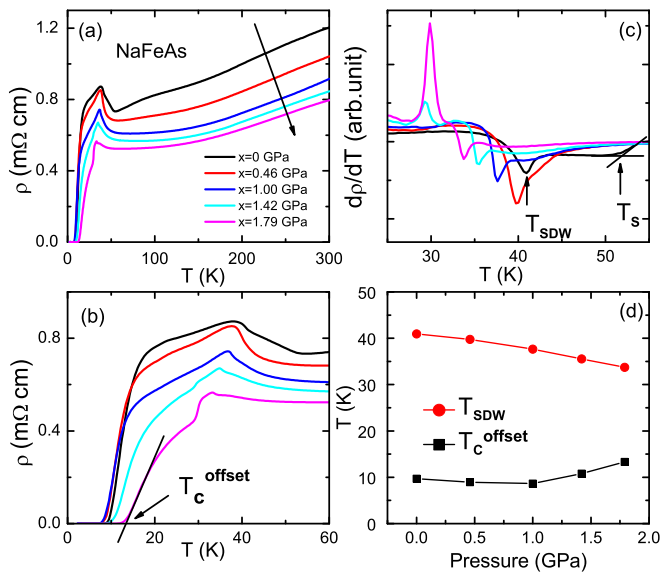}
\caption{(color online). (a): Temperature dependence of in plane
resistivity for NaFeAs under different pressures. Arrow indicates
the direction of the increasing pressure. (b): Expanded plot of the
temperature dependence of resistivity under various pressure around
$T_{\rm c}$. (c): The derivative of the in-plane resistivity
d$\rho$/d$\emph{T}$ and the criteria by which we infer $T_{\rm s}$
and $T_{\rm SDW}$. (d): $T_C$ and $T_{SDW}$ as a function of
pressure for the parent compound NaFeAs.} \label{fig1}
\end{figure}

\section{RESULTS AND DISCUSSION}

Fig. 1(a) shows the temperature dependence of resistivity for NaFeAs
under different pressures. Two anomalies in the resistivity curve
are observed at 51 K and 41 K under ambient pressure, which are
consistent with previous reports.\cite{ChenGF} These anomalies have
been proved by neutron scattering experiment to arise from the
structural and SDW transition, respectively.\cite{DaiPC} With
increasing pressure the anomalies corresponding to the SDW transition is
gradually suppressed to lower temperature, whereas the structural
transition quickly becomes undetectable. The suppression of the
resistive anomalies can also be seen in the derivative of
resistivity shown in Fig.1(c). We use the same criteria to infer the
structural and SDW transitions from the resistivity as
described in Ref.17, which has been confirmed by specific heat and
magnetic susceptibility measurements.\cite{WangAF} As shown in
Fig.1(b), the superconducting transition is rather broad, and the
three transitions take place in a narrow temperature range, thus it
is difficult to define the $T_{\rm c}^{\rm onset}$. We use the
criterion $T_{\rm c}^{\rm offset}$ to describe the superconducting
transition temperature in this paper, the definition of which is
shown in Fig.1(b). With the applied pressure increasing, $T_{\rm
c}^{\rm offset}$ firstly decreases slightly, then increases quickly
with pressure higher than 1 GPa. The highest $T_{\rm c}^{\rm
offset}$ we can achieve is 11.9 K at P = 1.79 GPa. The data of
NaFeAs under pressure are summarized in Fig.1(d). The phase diagram
T($P$) is similar to that of underdoped
Ba(Fe$_{1-x}$Co$_x$)$_2$As$_2$\cite{Ahilan}, in which $T_{\rm SDW}$
is suppressed gradually while superconductivity is enhanced by
applied pressure.

\begin{figure}[ht]
\centering
\includegraphics[width=0.49\textwidth]{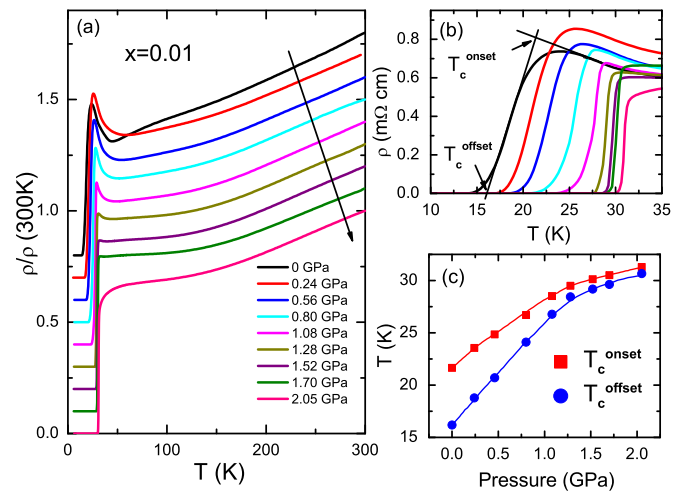}
\caption{(color online). (a): Temperature dependence of the
resistivity under different pressures up to 2.05 GPa for
NaFe$_{0.99}$Co$_{0.01}$As, normalized to the room temperature
value. Each subsequent data set is shifted downward by 0.1 for
clarity. (b): Enlargement of the low temperature resistivity and the
criteria used to determine the onset and offset temperatures for the
superconducting transitions. (c):  $T_C$ as a function of pressure
for the underdoped crystal NaFe$_{0.99}$Co$_{0.01}$As.} \label{fig2}
\end{figure}

For the underdoped sample NaFe$_{0.99}$Co$_{0.01}$As with $T_{\rm
c}^{\rm offset}$$\sim$16K, the kinks in resistivity curves associate
with the structural and SDW transition are distinct at ambient
pressure. Once the external pressure is applied, the kinks quickly
become obscure and then indistinguishable, similar to the case in
doped 122-system.\cite{Ahilan} As shown in Fig.2(a), the
low-temperature resistive upturn corresponding to the structural
and/or magnetic transition are progressively suppressed, and
ultimately vanished at P = 2.05 GPa, at which the highest
superconducting transition temperature about 30.7 K is obtained. The
criteria used to determine the onset and offset temperature of
superconducting transition are shown in Fig 2(b). Since the onset
temperature of superconductivity is ambiguous in
NaFe$_{1-x}$Co$_x$As, we write $T_{\rm c}$ for $T_{\rm c}^{\rm
offset}$ for convenience hereafter. As reported in underdoped
Ba(Fe$_{1-x}$Co$_x$)$_2$As$_2$, the critical pressure at which the
high temperature transition disappears coincides rather well with
the pressure at which $T_{\rm c}$ is highest and the superconducting
transition is narrowest. \cite{Colombier} The pressure coefficient
d$T_{\rm c}$/d$P$ is 9.6 K/GPa below 1.28 GPa, and the pressure
coefficient between ambient and the pressure at which $T_{\rm c}$
reaches its maximum is 7.06 K/GPa, even larger than the pressure
coefficient of FeSe (3.2 K GPa$^{-1}$).\cite{FeSepressure} The
pressure effect coefficient based on the $T_{\rm c}^{\rm onset}$ is
about 4.7 K/GPa which is still relatively large in iron pnictides.
In the phase diagram shown in figure 2(d), it is obvious that the
superconducting transition width become narrower with increasing the
pressure. The sharp superconducting transition observed at 2.05 GPa
indicates that the pressure condition is still hydrostatic.

\begin{figure}[ht]
\centering
\includegraphics[width=0.49\textwidth]{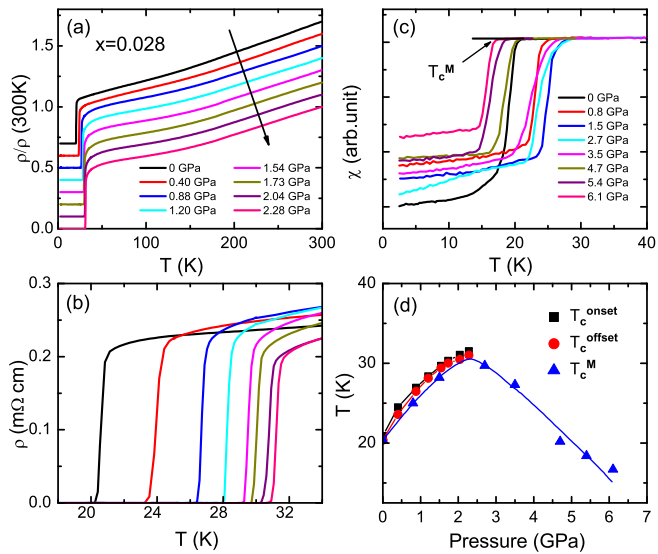}
\caption{(color online). (a) Temperature dependence of in plane
resistivity for NaFe$_{0.972}$Co$_{0.028}$As under various
pressures. Data are shown normalized by room temperature
resistivity, successive data sets are offset vertically by 0.1 for
clarity. (b) The data in panel (a) is plotted in low temperature
range for clarity. (c): The plot of Magnetic susceptibility as a
function of temperature in zero-field cooled measurements up to 6.1
GPa. (d): Evolution of $T_{\rm c}$ determined by resistivity and
susceptibility measurements with the applied pressure.} \label{fig3}
\end{figure}

For the optimally doped sample NaFe$_{0.972}$Co$_{0.028}$As, the
anomalies associated with the structural or SDW transition are
suppressed completely by Co doping. As shown in Fig.3(a), $T_{\rm
c}$ measured by in-plane resistivity increases monotonously from
20.4 K to 31.0 K with increasing the applied pressure from zero to
2.28 GPa. The pressure coefficient is 4.67 K/GPa, much larger than 1
K/GPa in optimally doped Ba(Fe$_{1-x}$Co$_x$)$_2$As$_2$\cite{Ahilan}
and comparable with 5 K/GPa in optimally doped
LaFeAsO$_{1-x}$F$_x$.\cite{Takahashi} In order to establish the
complete superconducting dome in the phase diagram. We carried out
the magnetic susceptibility measurement using the diamond anvil cell
(DAC) technology. Pressures up to 6.1 GPa were applied on
NaFe$_{0.972}$Co$_{0.028}$As single crystal. The $T_{\rm c}^{\rm M}$
values under various pressures are determined from the beginning of
deviation from the extrapolated line of the normal state $M$-$T$
curve as shown in Fig.3(c). Fig.3(d) displays the $T$($P$) phase
diagram based on the resistivity and magnetic susceptibility
measurements. The $T_{\rm c}^{\rm M}$ measured by DAC technology
initially increases monotonously, and begins to decrease when
pressure is higher than 2.3 GPa. The behavior of transition
temperature obtained by resistivity and magnetic susceptibility are
highly consistent with each other. The highest transition
temperature obtained by our measurement is 31.0 K, where the
transition width is 0.5 K which is considerably sharp. The $T_{\rm
c}$ = 31.0 K presented here is the highest in the 111 system up to
now.

\begin{figure}[ht]
\centering
\includegraphics[width=0.49\textwidth]{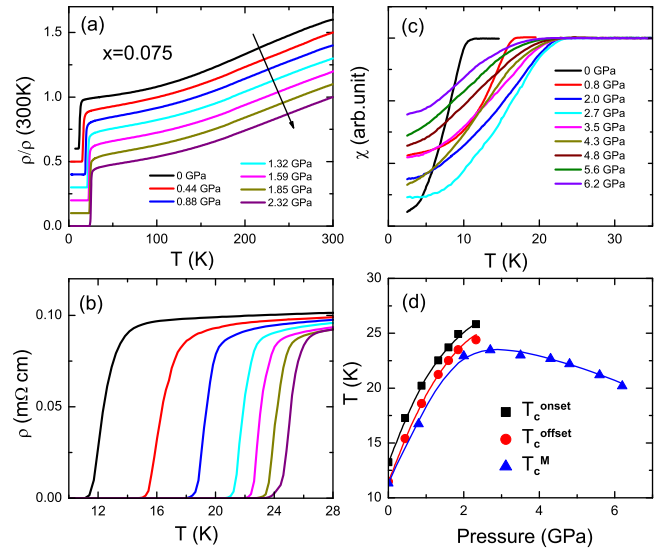}
\caption{(color online). (a): Temperature dependence of in plane
resistivity for NaFe$_{0.925}$Co$_{0.075}$As under various
pressures, Successive data sets are offset vertically by 0.1 for
clarity. (b): The same data shown in (a) around the superconducting
transition. (c): Temperature dependence of magnetic susceptibility
under different pressures for NaFe$_{0.925}$Co$_{0.075}$As. (d): The
$T_C$ obtained from resistivity and susceptibility measurements as a
function of pressure.} \label{fig4}
\end{figure}

Fig.4(a) and Fig.4(c) display the pressure dependence of the
in-plane resistivity and magnetic susceptibility of overdoped sample
NaFe$_{0.925}$Co$_{0.075}$As, respectively. The superconducting
transition temperature of this overdoped sample is 11.5 K at ambient
pressure. Similar to the case in optimally doped
NaFe$_{0.972}$Co$_{0.028}$As, the $T_{\rm c}$ increases monotonously
up to 24.5 K with increasing the applied pressure to 2.32 GPa.
Domelike shape of $T_{\rm c}$($P$) was revealed by magnetic
susceptibility measurement, from which we can infer that the highest
superconducting transition temperature in
NaFe$_{0.925}$Co$_{0.075}$As is about 24.5 K with the uncertainty
less than 1 K. A large enhancement of $T_{\rm c}$ by 13 K, which is
comparable to those in underdoped and optimally doped samples, is
still exist in this overdoped composition. The pressure coefficient
of NaFe$_{0.925}$Co$_{0.075}$As is 5.57 K/GPa, even higher than that
of the optimally doped sample. The large pressure coefficient of
overdoped NaFe$_{0.925}$Co$_{0.075}$As is obviously different from
those in other iron-pnictide superconductors, which are rather small
or change their sign from positive to negative in the overdoped
regime in the phase diagram.\cite{Ahilan,SmOFeAs,Torikachvili,KSr}

\begin{figure}[ht]
\centering
\includegraphics[width=0.49\textwidth]{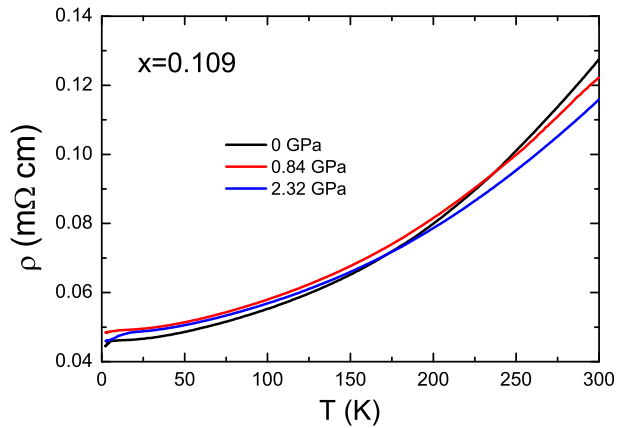}
\caption{(color online). Temperature dependence of in plane
resistivity for heavily overdoped non-superconducting crystal
NaFe$_{0.891}$Co$_{0.109}$As under various pressures.} \label{fig5}
\end{figure}

When external pressure was applied on the extremely overdoped sample
NaFe$_{0.891}$Co$_{0.109}$As which shows no superconductivity down
to 2 K at ambient pressure, we cannot observe the pressure-induced
superconductivity with applied pressure up to 2.32 GPa. Besides,
both the magnitude and the behavior of the resistivity do not change
much with the applied pressure.

\begin{figure}[ht]
\centering
\includegraphics[width=0.49\textwidth]{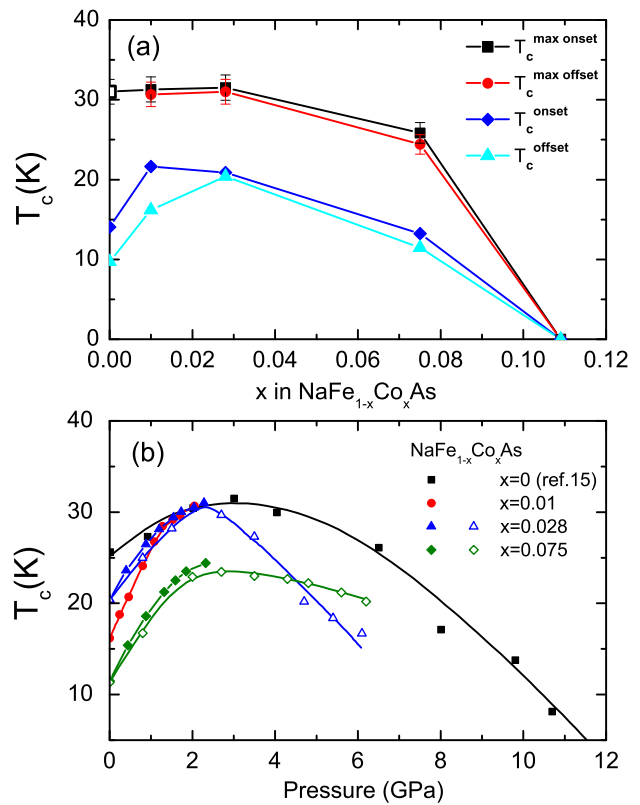}
\caption{(color online). (a): Comparison of $T_{\rm c}$ at ambient
pressure and the maximum $T_{\rm c}$ achieved under applied pressure
at various Co concentrations. The open square represent the $T_{\rm
c}^{\rm onset}$ reported by Zhang $et$ $al.$\cite{JinCQ} (b):
$T$($P$) phase diagram of NaFe$_{1-x}$Co$_x$As with different doping
levels. Open and filled symbols represent data obtained from
susceptibility and resistivity measurements,
respectively.}\label{fig.6}
\end{figure}

The effect of applied pressure on the superconducting transition
temperature of NaFe$_{1-x}$Co$_x$As is illustrated in Fig.6(a),
where the maximum $T_{\rm c}$ under pressure and $T_{\rm c}$ at
ambient pressure are plotted as a function of doping level $x$.
Because the maximum transition temperature has not been reached in
undoped NaFeAs, we use the maximum value reported by Zhang $et$
$al$. for this composition.\cite{JinCQ} Large positive pressure
coefficient d$T_{\rm c}$/d$P$ is observed in all the superconducting
compositions, even in the overdoped regime. The maximum transition
temperatures in NaFeAs, NaFe$_{0.99}$Co$_{0.01}$As and
NaFe$_{0.972}$Co$_{0.028}$As are all around 31 K. $T$($P$) phase
diagram of NaFe$_{1-x}$Co$_x$As system for various $x$ is shown in
Fig.6(b). It is obvious that the maximum transition temperature in
undoped, underdoped and optimally doped sample is strikingly the
same. These results indicate that there is a universal maximum
transition temperature in NaFe$_{1-x}$Co$_x$As of about 31 K, which
can be achieved by applying a critical pressure of P = 2-3 GPa.

For the undoped and underdoped samples, applied external pressure
suppresses the structural and SDW transitions, and enhances
superconductivity simultaneously. This behavior is similar to the
pressure effect in LaFeAsO$_{1-x}$F$_x$ \cite{Takahashi} and
122-systems.\cite{Alireza,SrFe2As2pressure,Ahilan,Torikachvili}
Although the $T_{\rm c}$ and normal state resistivity behavior
evolve systematically with Co doping, the maximum transition
temperature enhancement and corresponding critical pressure is
nearly the same in all the superconducting samples. These properties
are different from the case in most of the iron based
superconductors that pressure effect is different in different
regions of electronic phase diagram. An identical maximum $T_{\rm
c}$ of 31 K under doping as well as pressure has also been reported
in BaFe$_2$(As$_{1-x}$P$_x$)$_2$.\cite{BaFe2As2P} In the P doped
Ba-122 system, phosphorous substitution could be regarded as
chemical pressure, which changes Fe-Pn distance and causes similar
effects on superconductivity to the physical pressure. Though the Co
substitution in NaFeAs is referred to as electron doping, different
from the replacement of As by P which is referred to as isovalent
substitution, it is likely that the pressure-induced enhancement of
$T_{\rm c}$ in NaFe$_{1-x}$Co$_x$As is also associated with the
optimization of the structural parameters of FeAs layers, including
the As-Fe-As bond angle and anion height.\cite{QLiu} One possible
reason for lower maximum transition temperature obtained in
overdoped NaFe$_{0.925}$Co$_{0.075}$As single crystal is that the
superconductivity is disturbed by the disorder or additional
scattering induced by excess cobalt doping. This phenomenon is
different from the case of overdoped LaFeAsO$_{1-x}$F$_x$, in which
the conducting layer is not affected by F doping and the highest
transition temperature acquired in optimal and overdoped samples are
almost the same.\cite{Takahashi}

The overdoped superconducting sample NaFe$_{0.925}$Co$_{0.075}$As
still has considerable positive pressure coefficient which is rare
in Fe-pnictide superconductors. However, when the pressure is
applied on the extremely overdoped non-superconducting sample, no
superconductivity induced by pressure can be observed. It has also
been reported that temperature linear dependent susceptibiltiy can
be observed in high temperatures for all the superconducting samples
and the breakdown of the temperature linear dependent susceptibility
in the overdoped region coinciding with the disappearance of
superconductivity.\cite{WangAF} These phenomena indicate that there
is a sudden change in the electronic structure between the
superconducting compositions and the heavily overdoped
non-superconducting phase. This conclusion is supported by the
experimental results of STM investigations\cite{WangYY} and the
angle-resolved photoemission spectroscopy (ARPES)
studies.\cite{ZSun} The STM study revealed that the high energy
$dI/dV$ spectra of superconducting NaFe$_{1-x}$Co$_x$As remain
nearly the same, whereas, the high energy spectrum suddenly started
to shift to the lower energy substantially for the sample with x =
0.109. The direct measurements of the electronic structure of
NaFe$_{1-x}$Co$_x$As by ARPES revealed that all the superconducting
NaFe$_{1-x}$Co$_x$As compounds have similar band structures and
small relative Fermi level shifts. However, the x = 0.109 compound
in the heavily overdoped regime shows a large Fermi level shifts
(about 100 meV) relative to the optimally doped compounds, and its
band structure is significantly changed as the hole-like bands
around the zone center disappears and an electron pocket appears
instead, which means the consequent Fermi surface consists of
electron pockets only in this sample. The drastic change in
electronic structure for the heavily overdoped nonsuperconducting
samples could explain the observed properties of pressure effect.\\

\section{Summary}

In conclusion, we have performed resistivity and magnetic
susceptibility measurements under various pressure on
NaFe$_{1-x}$Co$_x$As (x=0, 0.01, 0.028, 0.075, 0.109) single
crystals. In the undoped and underdoped compounds, structural and
SDW transitions are gradually suppressed while superconductivity is
enhanced by applied external pressure. A universal maximum
transition temperature of about 31 K under external pressure is
observed in underdoped and optimally doped
NaFe$_{1-x}$Co$_x$As. The superconducting transition temperature of
NaFe$_{1-x}$Co$_x$As is strongly enhanced in the whole
superconducting regime of the phase diagram, and the pressure effect
is considerably large compared to other iron pnictides. The large
positive pressure coefficient in the optimally and overdoped region is
different from that in Ba(Fe$_{1-x}$Co$_x$)$_2$As$_2$, and
disappears simultaneously with the superconductivity in the phase
diagram. These results could be explained as originated from the
similarity of electron structures within the superconducting
dome, and a drastic change of the electron structures between the
superconducting overdoped regime and the non-superconducting heavily
overdoped regime, which correspond to the conclusions of STM and
ARPES measurements.

\section*{Acknowledgements}This work is supported by the National Natural Science Foundation of China
(Grant No. 11190021 and No. 51021091), the National Basic Research
Program of China (973 Program, Grant No. 2012CB922002 and No.
2011CB00101) and the Chinese Academy of Sciences.

\end{document}